\documentclass[aps,twocolumn,times,pre,10pt]{revtex4-1}

\usepackage{epsfig}
\usepackage{graphicx}
\usepackage{epstopdf}
\usepackage{amssymb}
\usepackage{amsmath}
\usepackage{amsfonts}
\usepackage{mathrsfs}
\usepackage{float}
\usepackage[usenames,dvipsnames]{xcolor}
\definecolor{webgreen}{rgb}{0,0.75,0}
\definecolor{webred}{rgb}{0.75,0,0}
\definecolor{webblue}{rgb}{0,0,0.75}
\definecolor{darkblue}{rgb}{0,0,0.7}
\definecolor{dunkelgrau}{rgb}{0.8,0.8,0.8}
\definecolor{lgray}{rgb}{0.95,0.95,0.95}
\definecolor{lgreen}{rgb}{0.95,1.00,0.90}
\definecolor{lblue}{rgb}{0.9,0.95,1.00}
\definecolor{lred}{rgb}{1.00,0.90,0.80}
\definecolor{shadecolor}{rgb}{1.00,0.92,0.82}

\usepackage[breaklinks]{hyperref}
\hypersetup{colorlinks=true, linkcolor=darkblue, citecolor=darkblue, filecolor=darkblue, urlcolor=darkblue}
\usepackage[hyphenbreaks]{breakurl}

\def\dj{d\kern-0.4em\char"16\kern-0.1em}
\def\Dj{\mbox{\raise0.3ex\hbox{-}\kern-0.4em D}}

\date{}
\begin{document}

\title{Influence of anharmonic convex interparticle potential and Shapiro steps in the opposite direction of driving force}

\author{Sonja Gombar$^{1}$, Petar Mali$^{1}$, Slobodan Rado\v sevi\' c$^{1}$, Jasmina Teki\' c$^2$, Milan Panti\' c$^{1}$, Milica Pavkov-Hrvojevi\' c$^{1}$}
\affiliation{$^1$ Department of Physics, Faculty of Sciences, University of Novi Sad,
Trg Dositeja Obradovi\' ca 4, 21000 Novi Sad, Serbia}
\affiliation{$^2$"Vin\v ca" Institute of Nuclear Sciences,
Laboratory for Theoretical and Condensed Matter Physics - 020,
University of Belgrade, PO Box 522, 11001 Belgrade, Serbia}

\date{\today}

\begin{abstract}
The response function and largest Lyapunov exponent analysis were applied to the driven overdamped Frenkel-Kontorova model with two types of anharmonic convex interparticle potentials. In both cases model reduces to a single particle model for integer values of winding number. It is shown that the mirror image of the amplitude dependence of critical depinning force and largest Lyapunov exponent observed recently in the standard Frenkel-Kontorova model [\href{https://www.sciencedirect.com/science/article/pii/S1007570416304142}{Commun. Nonlinear Sci. Numer. Simul. \textbf{47}, 100 (2017)}] is not retained generally. Behaviour of systems with relatively strong interparticle force was examined and evidence for the appearance of mode-locking phenomenon in both directions of particles' motion is presented. 
\end{abstract}

 \maketitle

\section{Introduction}

In the past few decades, synchronization phenomenon has become a subject of particular interest. Although it can be detected in many systems, the focus is on examination of dynamical mode-locking, which emerges when systems with a characteristic internal frequency are subjected to an external periodic drive \cite{synch,koku}. For instance, Shapiro steps (i.e., step-like macroscopic response) are experimentally observed in various colloidal systems \cite{e1,e2,c1,c2,c3}, charge-density wave systems  \cite{thorn1,e3,thorn}, vortex lattices \cite{e4,e+,e++}, Josephson junction arrays \cite{e5,e6,e7,e8,e9} and others. Regardless of the response function type (average velocity as a function of $F_{\textrm{dc}}$ force, $IV$ characteristics), both pinning-sliding transition and Shapiro steps can be recognized. In addition, special attention is given to the critical force \cite{fut1} and Shapiro steps width dependence on value of the ac force amplitude in these systems, which has proven to be Bessel-like \cite{e3,thorn,grun,hebb,sell}. 

A model which is used to successfully capture such behaviour in many condensed matter systems, micro- and nanotechnologies in the past few decades is the Frenkel-Kontorova model (FK model) driven by periodic forces \cite{brkiv,liwa,petar}. The FK model is commonly used to describe the dynamics of Josephson junction arrays \cite{usti,black,n1,n2}, DNA chains \cite{chafer}, charge-density waves \cite{flomazo,grun,midi}, incommensurate phases in dielectrics, dynamics of domain walls in ferromagnetic and antiferromagnetic chains \cite{brkiv}, etc. The one-dimensional Frenkel-Kontorova model describes chain of identical particles which are coupled to their nearest neighbours and subjected to a substrate potential. In a special case of standard FK model, interparticle potential is harmonic and substrate potential is sinusoidal \cite{flomazo}. A special attention is given to theoretical research of the dissipative FK model driven by dc and ac forces in both overdamped \cite{1,2,3,4} and underdamped \cite{5,6,7} regimes nowadays. 

From the theoretical point of view, several interesting results were found for the standard dissipative FK model driven by external periodic forces \cite{flomazo}. For example, width of the first harmonic Shapiro step and the critical force exhibit Bessel-like behaviour and the maxima of one function correspond to the minima of the other \cite{1}. Also, the standard FK model cannot be used for modeling phenomena related to subharmonic steps since subharmonic steps do not exist in the commensurate structures with integer values of winding number \cite{dod,dod1}, while for the non-integer values their size is too small \cite{flomazo}. On the other hand, dissipative ac+dc driven FK model with various forms of deformable substrate potentials was very successful in describing most of the phenomena related to the Shapiro steps such as the origin of subharmonic steps, their amplitude and frequency dependence \cite{2,8}. Since detection of subharmonic steps can be particularly difficult, the largest Lyapunov exponent (largest LE) analysis is often used instead of the response function as it represents a more sensitive way to detect subharmonic steps \cite{3}.

Most of the research done so far regarded the harmonic type of interparticle potentials and different forms of periodic substrate potentials with a periodic driving force (see \cite{1} for example), but a more general type of interparticle potential has not been widely examined yet. However, some properties of the FK model with non-convex interparticle potentials \cite{noncon,noncon2} and dynamics of the FK systems with Morse \cite{anharm} and Toda interparticle potentials \cite{anharm2} have been investigated in the last few decades. An anharmonic interparticle potential (AIP) form is present in the model description of Josephson junctions \cite{shuk} and many other systems \cite{brakiv}. Consequently, it is of significant importance to investigate the effect of AIPs on dynamics in the FK systems driven by external periodic force.

The purpose of this paper is to demonstrate how a change of harmonic interparticle potential affects the model dynamics using both response function and largest LE analysis and compare it to the previously well-investigated standard FK model case. Since in the one-dimensional overdamped FK model with convex interparticle potential the particles move following Middleton's no passing rule (i.e., they preserve their order during motion), the system possesses the property of asymptotic uniqueness. This means that, in the limit of long times, the average velocities of the particular commensurate structure approach a unique solution \cite{midl,slijep}. Therefore, the overdamped FK models with sinusoidal substrate potential and convex AIPs that depend only on distance between the nearest neighbours are studied in this paper. Two types of AIPs are taken into account and their influence on the depinning force, size of the steps and their amplitude dependence is examined.  

The paper is organized as follows. A brief description of the model is given in Section II, while the results are presented in Section III. The paper ends with discussion of the results in Section IV.

\section{Model}

Total potential energy of the standard FK model is given by \cite{flomazo}:
\begin{equation}
H=\sum_{j}\left(V(u_{j})+W(u_{j+1}-u_{j})\right), \hspace{2mm} j=1,2,...,N,
\end{equation}
where $N$ is the number of particles, $V(u_{j})$ is sinusoidal substrate potential of the form:
\begin{equation}
V(u_{j})=\frac{K}{(2\pi)^{2}}\left(1-\cos(2\pi u_{j})\right), \label{potenc}
\end{equation}
with strength of the periodic substrate potential $K$, and $W(u_{j+1}-u_{j})$ is pairwise harmonic interparticle potential, which is a function of distance between the nearest neighbours:
\begin{equation}
W(u_{j+1}-u_{j})=\frac{1}{2}(u_{j+1}-u_{j})^{2}.
\end{equation}
This form of the interparticle potential is the same as the one from \cite{obr}. 
\par In the present paper the overdamped FK model subjected to the influence of both ac and dc external force is considered. The corresponding equations of motion are:
\begin{equation}
\dot{u}_{j}=\nabla^2 u_j +\frac{K}{2\pi}\sin(2\pi u_{j})+F(t), \hspace{2mm} j=1,2,...,N \label{1}
\end{equation}
with $\nabla^2$ being the lattice Laplacian \cite{midl,saad}:
\begin{equation}
\nabla^2 u_j = u_{j+1} + u_{j-1} - 2 u_{j},
\end{equation}
and external force $F$ has the following form:
\begin{equation}
F(t)=F_{\textrm{dc}}+F_{\textrm{ac}}\cos(2\pi \nu_{0}t). \label{sila}
\end{equation}
$F_{\textrm{dc}}$ is the dc force, $F_{\textrm{ac}}$ is amplitude of the ac force and $\nu_{0}=\frac{1}{T}$ is frequency of the ac force with $T$ being its period. 

Due to the competition between two frequency scales (one of the external periodic force and the other of the motion associated with the periodic substrate potential), the dynamics of this model is characterized by the appearance of Shapiro steps. These steps correspond to the resonant solutions of \eqref{1}. If $\lbrace u_{j}(t)\rbrace$ is the solution of \eqref{1} with the initial condition $\lbrace u_{j}(t_{0})\rbrace$, then:
\begin{equation}
\sigma_{r,m,s}\lbrace u_{j}(t)\rbrace=\lbrace u_{j+r}(t-\frac{s}{\nu_{0}})+m\rbrace \label{2}
\end{equation}
is also a solution of the same equations corresponding to the initial condition $\sigma_{r,m,s}\lbrace u_{j}(t_{0})\rbrace$, where $r$, $m$ and $s$ are arbitrary integers. The index $r$ defines the relabeling of particles, while $m$ and $s$ determine space and time translations, respectively. Validity of symmetry transformation \eqref{2} can easily be verified by a simple substitution:
\begin{widetext}
\begin{align}
\dot{u}_{j+r}(t-\frac{s}{\nu_{0}})=u_{j+r+1}(t-\frac{s}{\nu_{0}})+u_{j+r-1}(t-\frac{s}{\nu_{0}})-2u_{j+r}(t-\frac{s}{\nu_{0}})-\frac{K}{2\pi}\sin\left(2\pi\left(u_{j+r}(t-\frac{s}{\nu_{0}})+m\right)\right)+F(t). \label{3}
\end{align} 
\end{widetext}
Since 
\begin{align}
\sin\left(2\pi\left(u_{j+r}(t-\frac{s}{\nu_{0}})+m\right)\right)=\sin\left(2\pi u_{j+r}(t-\frac{s}{\nu_{0}})\right) \nonumber
\end{align}
 and $F(t)$ is a function periodic in time, with a period $T=\frac{1}{\nu_{0}}$, equations \eqref{1} and \eqref{3} are equivalent and \eqref{2} is another solution of \eqref{1}. The solution is resonant if it is invariant under symmetry operation \eqref{2}. The corresponding average velocity is then easily evaluated:
\begin{eqnarray}
\bar{v}&=&\frac{1}{sT}\frac{1}{N-M}\sum_{j=M}^{N-1}\left(u_{j}(t+sT)-u_{j}(t)\right) \nonumber \\
&=&\frac{\nu_{0}}{s}\frac{1}{N-M}\sum_{j=M}^{N-1}\left(u_{j+r}(t)+m-u_{j}(t)\right)
\end{eqnarray} 
and the final form is well-known \cite{petar,flomazo,3,falo}:
\begin{equation}
\frac{\bar{v}}{\nu_{0}}=\frac{r\omega+m}{s}, \label{4}
\end{equation}
where $\omega$ is the winding number, i.e. the average distance between particles in a configuration. For example, $\omega=1/2$ corresponds to the configuration in which there are, on average, two particles per minimum of the substrate potential. The Shapiro steps are called harmonic if $s=1$, whereas they are labeled as subharmonic if $s>1$.

Although equivalence of some non-convex and convex models was shown previously \cite{sagri}, only convex interparticle potentials, $W''(u)>0$, are considered in this paper. First off, it is of particular interest to establish how anharmonic terms affect the model dynamics. Thus, quartic polynomial AIP was used:
\begin{eqnarray}
W(u_{j+1}-u_{j}) &=&\frac{g}{2}(u_{j+1}-u_{j})^{2}+\frac{h}{3}(u_{j+1}-u_{j})^{3} \nonumber \\
&+&\frac{f}{4}(u_{j+1}-u_{j})^{4} \label{5}
\end{eqnarray} 
where constants $g$, $h$ and $f$ are chosen conveniently in order to obtain convex interparticle potential. 
The corresponding equations of motion are:
\begin{equation}
\dot{u}_{j}=g \nabla^2 u_j-h(\nabla^2 u_j)^2+f(\nabla^2 u_j)^{3}+\frac{K}{2\pi}\sin(2\pi u_{j})+F(t), \label{*}  
\end{equation}
where $F(t)$ is given by \eqref{sila} and $j=1,2,...,N$. Parameters $g$, $h$ and $f$ will be varied and their influence on the model dynamics examined in the following section. 

The second type of AIP examined in this paper is given by exponential form:
\begin{equation}
W(u_{j+1}-u_{j})=e^{-(u_{j+1}-u_{j})} \label{6}
\end{equation}
with the corresponding equations of motion:
\begin{align}
\dot{u}_{j}=e^{-(u_{j}-u_{j-1})}-e^{-(u_{j+1}-u_{j})}&+\frac{K}{2\pi}\sin(2\pi u_{j})+F(t), \nonumber \\ j=1,2,&...,N \label{**}
\end{align}

Since the convex AIPs \eqref{5} and \eqref{6} are functions of distance between the nearest neighbours and independent of time, the symmetry of the solution is the same as in \eqref{2} and the resonant velocities are given by \eqref{4}. 

The response functions were obtained from the equations of motion \eqref{*} and \eqref{**}, which had been integrated using the fourth order Runge-Kutta method with periodic boundary conditions. The dc force was varied adiabatically with the step $\Delta F_{\textrm{dc}}=10^{-5}$. The largest LE $\lambda$ was calculated according to the algorithm given in details in \cite{spro}. In order to calculate the largest LE we choose a perturbed point $\tilde{u}_{j}$ given by:
\begin{equation}
\tilde{u}_{j}(t_{\textrm{ss}})=u_{j}(t_{\textrm{ss}})\pm \sqrt{\frac{d_{0}^{2}}{N}},
\end{equation}
where $t_{\textrm{ss}}$ is the necessary time for system to reach the steady-state and $d_{0}=10^{-7}$ is the small parameter that defines perturbation of the initial configuration. The plus and minus sign in the equation appear with the equal probability so that projecting is always done onto the subspace dominated by the largest LE \cite{oda}. In numerical calculations one will always deal with finite-time LEs. For large values of $t_{\textrm{ss}}$ the finite-time LEs converge to their asymptotic limit. Time evolution of full Lyapunov spectrum of driven FK model is discussed in \cite{doda}. In the case when it is not specified in the description of a figure, we used $t_{\textrm{ss}}=300\nu_{0}^{-1}$.

\section{Results}

In this section, the study of FK model with two types of convex AIPs is presented. Both response function and largest LE analysis were used.

\subsection{Subharmonic steps and the largest Lyapunov exponent analysis}

\begin{figure}[]
\includegraphics[width=\columnwidth]{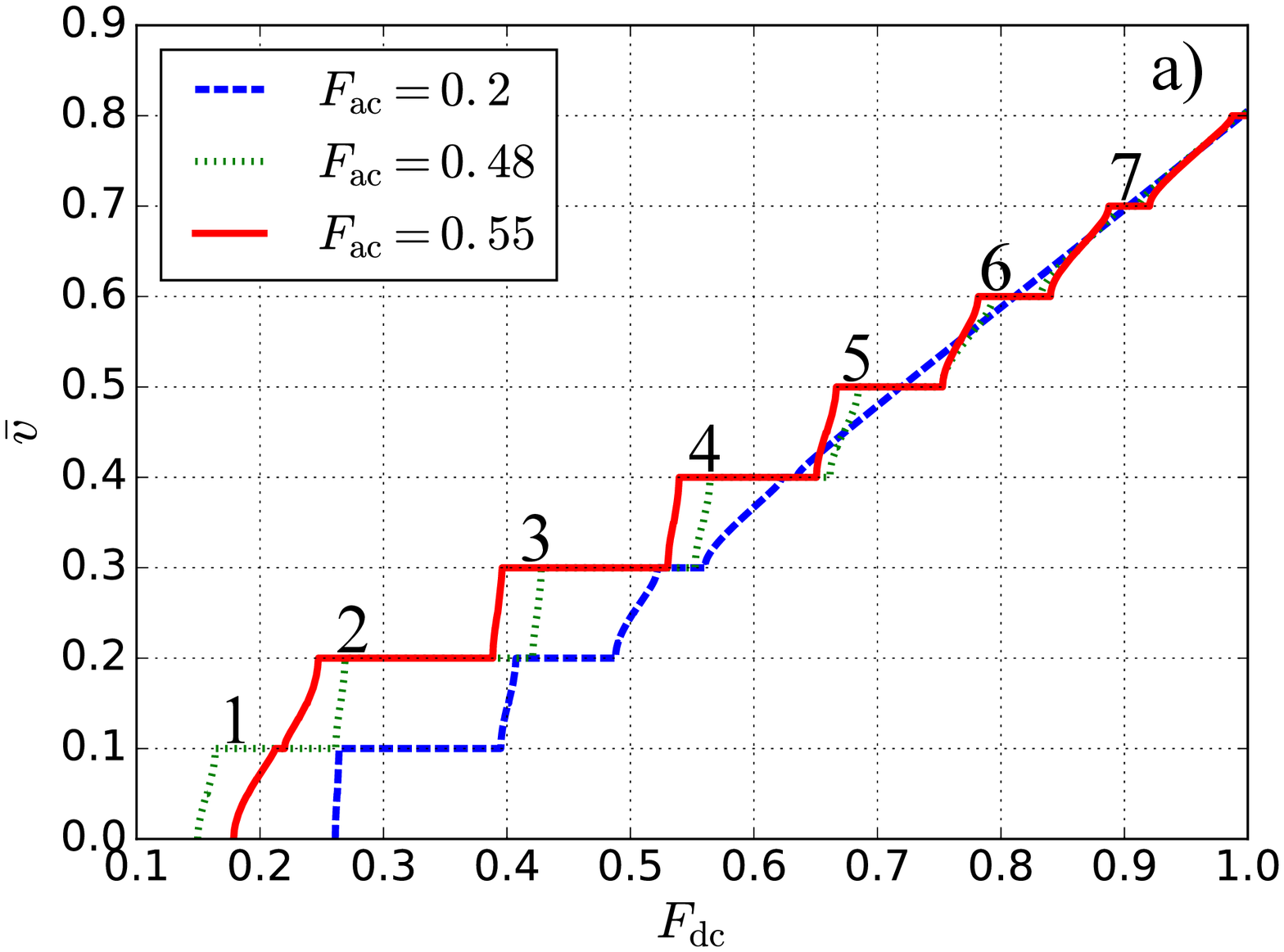} \includegraphics[width=\columnwidth]{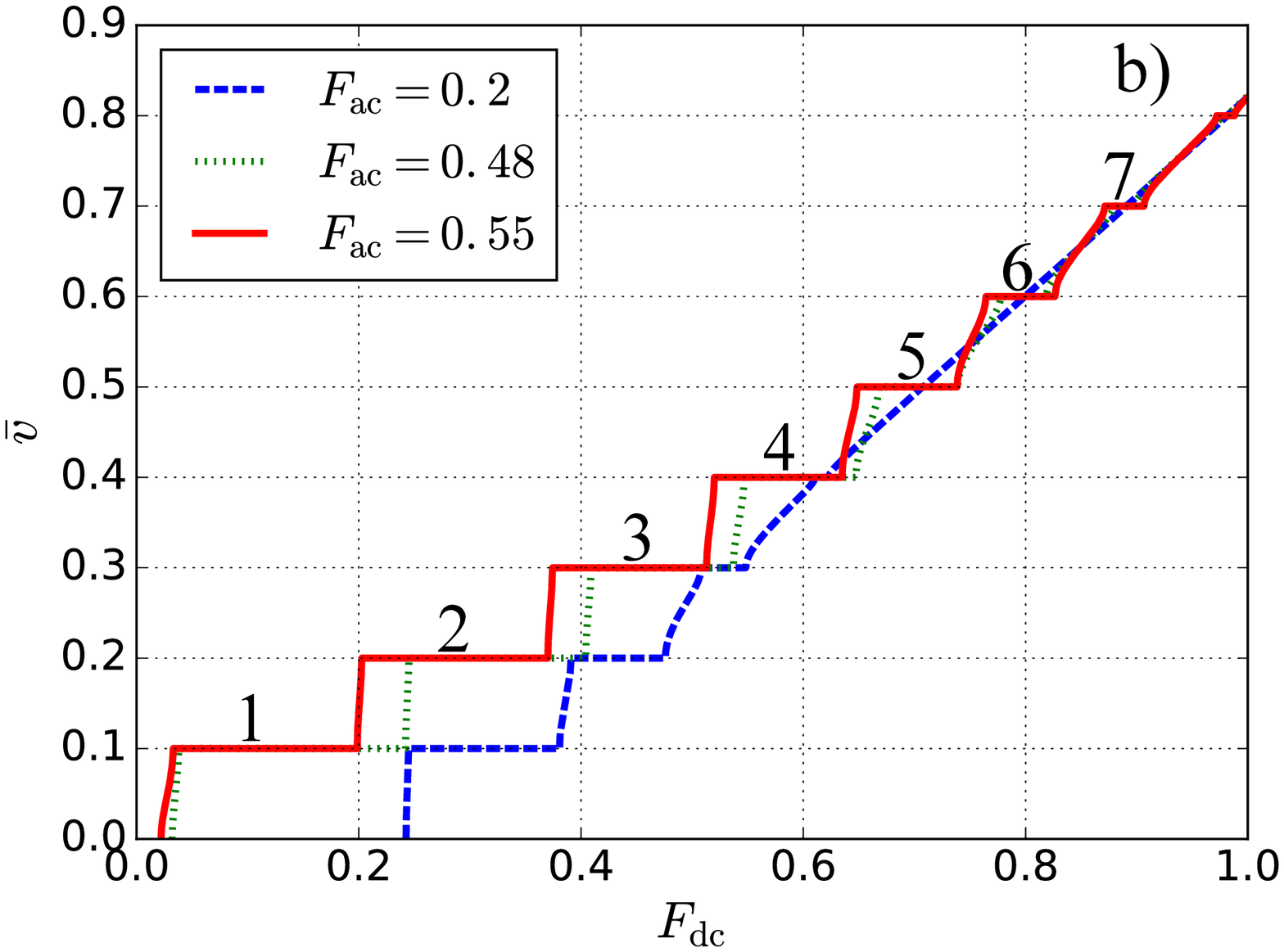}
\centering
\caption{Average velocity as a function of driving force for two AIPs: quartic polynomial \eqref{5} with $g,h,f=1$ (a)) and exponential interparticle potentials \eqref{6} (b)) and three values of the ac force amplitude. The chosen set of parameters is: $\omega=\frac{1}{2}$, $K=4.0$ and $\nu_{0}=0.2$. The numbers mark harmonic steps.}
\label{fig:f1}
\end{figure}

The response functions $\bar{v}=\bar{v}(F_\textrm{dc})$ corresponding to the equations of motion \eqref{*} and \eqref{**} are presented in Figure \ref{fig:f1}. Since the same symmetry \eqref{2} accounts for two chosen types of AIPs, the same resonant velocities \eqref{4} are expected to be observed in the $\bar{v}=\bar{v}(F_\textrm{dc})$ plot. Namely, for $\omega=\frac{1}{q}$, one can always choose $m=0$ (see \cite{flomazo}), which simplifies the expression \eqref{4}:
\begin{equation}
\bar{v}=\frac{r\omega}{s}\nu_{0}. \label{8}
\end{equation}

\begin{figure}[]
\includegraphics[scale=0.35]{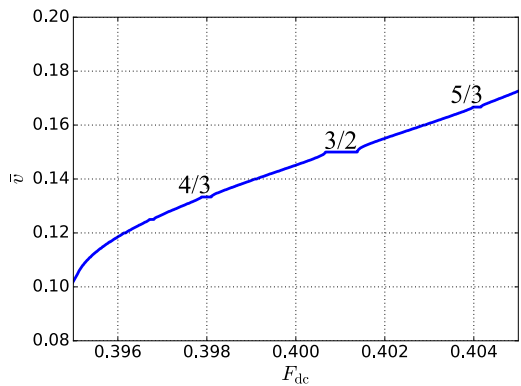}
\centering 
\caption{Zoomed segment of Figure \ref{fig:f1} a) for $F_{\textrm{ac}}=0.2$. The numbers mark subharmonic steps.}
\label{fig:f2}
\end{figure}

Given the equation \eqref{8}, all the Shapiro steps observed in Figure \ref{fig:f1} are harmonic. The response functions are somewhat similar to the ones of the standard FK model \cite{falo}. In addition, high resolution analysis reveals the subharmonic steps, which is shown in case of the quartic AIP in Figure \ref{fig:f2}. The same is valid in case of the exponential AIP.  

Further, we have verified that for $\omega=1$ subharmonic steps are not detected and that the response functions corresponding to equations \eqref{*} and \eqref{**} coincide with the standard FK model case for the same set of parameters. It was shown previously that for the standard harmonic interparticle potential and integer values of winding number, the model reduces to a single particle model \cite{flomazo}. The examination of response functions for multiple sets of parameters and various integer values of winding number has shown that subharmonic steps do not exist even in the case of AIPs considered in this paper and thus new degrees of freedom are not added to the system. 

In spite of analogous behaviour, there are some deviations from the standard model in two cases considered in this paper. The values of both critical force $F_{\textrm{c}}$ and size of Shapiro steps vary with the change of interparticle potential form. Furthermore, the values of critical force are a bit larger than the one obtained in purely harmonic case \cite{falo,1}. This may lead to the conclusion that the anharmonic form of interparticle potential affects particles' tendency to leave their positions pinned by the periodic substrate potential and, for the given set of parameters, they need to be stimulated by a larger value of dc drive. 

\begin{figure}[H]
\includegraphics[width=\columnwidth]{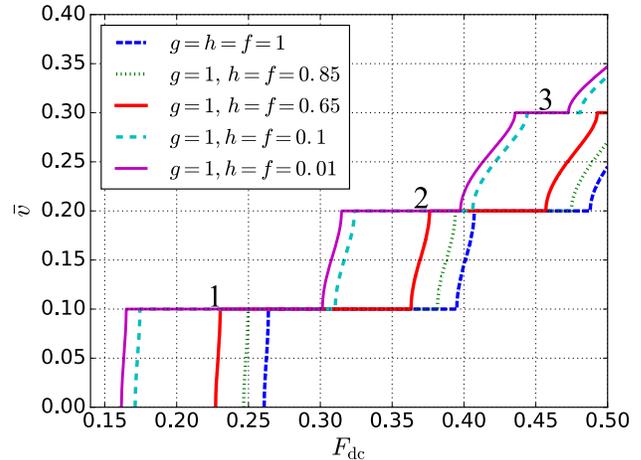} 
\centering
\caption{Average velocity as a function of driving force in case of quartic polynomial interparticle potential \eqref{5} for $\omega=\frac{1}{2}$, $F_{\textrm{ac}}=0.2$, $K=4.0$, $\nu_{0}=0.2$ and five sets of parameters $g$, $h$ and $f$. The numbers mark harmonic steps.}
\label{fig:f3} 
\end{figure}

Since the quartic polynomial AIP \eqref{5} reduces to the standard harmonic form for $g\to 1$, $h,f\to 0$, plot in Figure \ref{fig:f1} a) is expected to turn into its standard FK form from \cite{falo} (the same applies for other ac force frequencies). This situation is presented in Figure \ref{fig:f3}. The critical depinning force tends towards its value in case of the standard FK model (approximately $F_{\textrm{c}}=1.6$ for given set of parameters \cite{falo,1}). Varying the values of this set of parameters does not influence widening of the subharmonic steps, as only harmonic steps are detected in Figure \ref{fig:f3}, and high resolution analysis is necessary in order to observe them. 

\begin{figure*}[ht]
\includegraphics[scale=0.65]{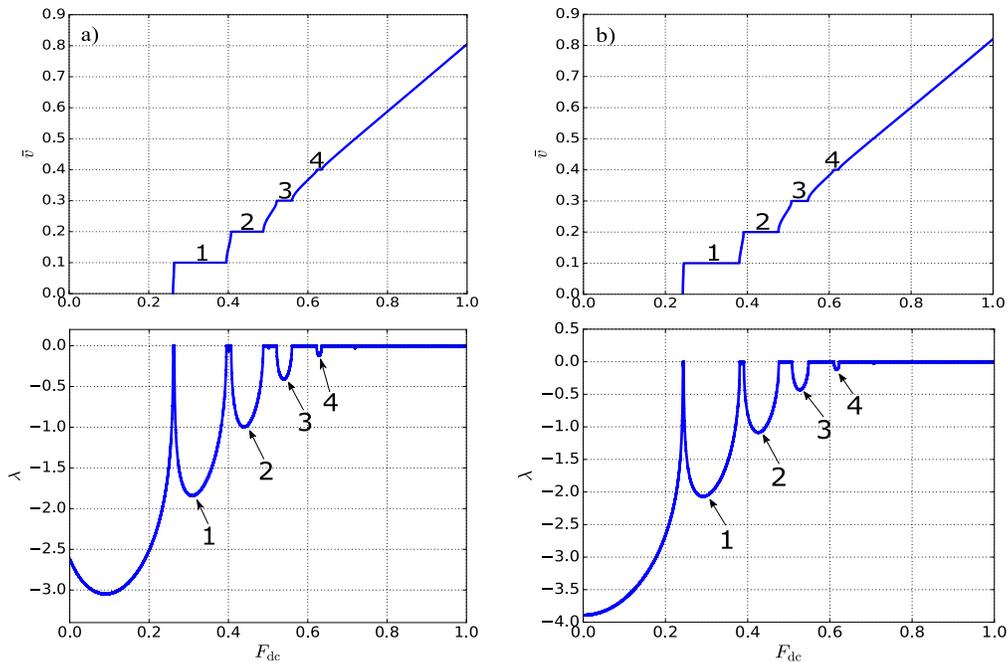}
\centering
\caption{Average velocity and largest LE as functions of driving force in case of two AIPS: a) quartic polynomial \eqref{5} and b) exponential interparticle potentials \eqref{6}. The chosen set of parameters is: $\omega=\frac{1}{2}$, $F_{\textrm{ac}}=0.2$, $K=4.0$, $\nu_{0}=0.2$ and $g=h=f=1$. The numbers mark harmonic steps.}
\label{fig:f4}
\end{figure*}

LEs determine the rate of separation of the infinitesimally close trajectories. Due to the fact that the systems of differential equations \eqref{*} and \eqref{**} both consist of $N$ first order equations, the corresponding Lyapunov spectrum has $N$ exponents \cite{doda}. The long-term linear stability of a given trajectory is characterized by the largest LE. The largest LE analysis is useful since it allows one to examine the appearance of chaos in a system \cite{hil,4,7}. If the largest LE is positive, the system is chaotic. It is well-known that chaotic behaviour has not been observed in the overdamped FK model with various types of substrate potentials \cite{oda}. However, the question remains whether taking the discussed AIPs into account would result in the appearance of chaos in the system. The results of comparison between the response function and largest LE analysis are given in Figure \ref{fig:f4}. 

\begin{figure}%[H]
\includegraphics[scale=0.32]{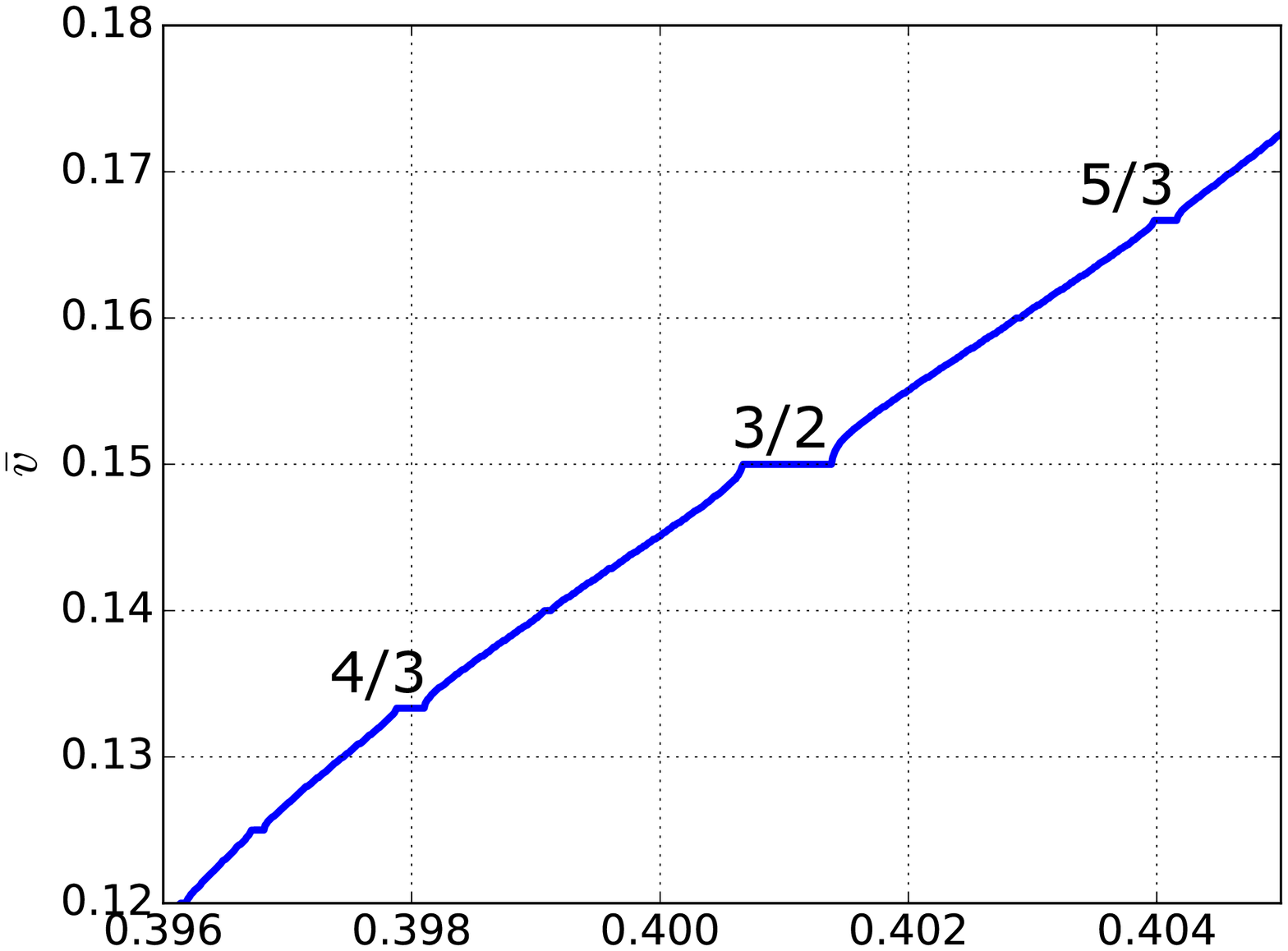}
\centering \\ \hspace{-3.1mm} \includegraphics[scale=0.414]{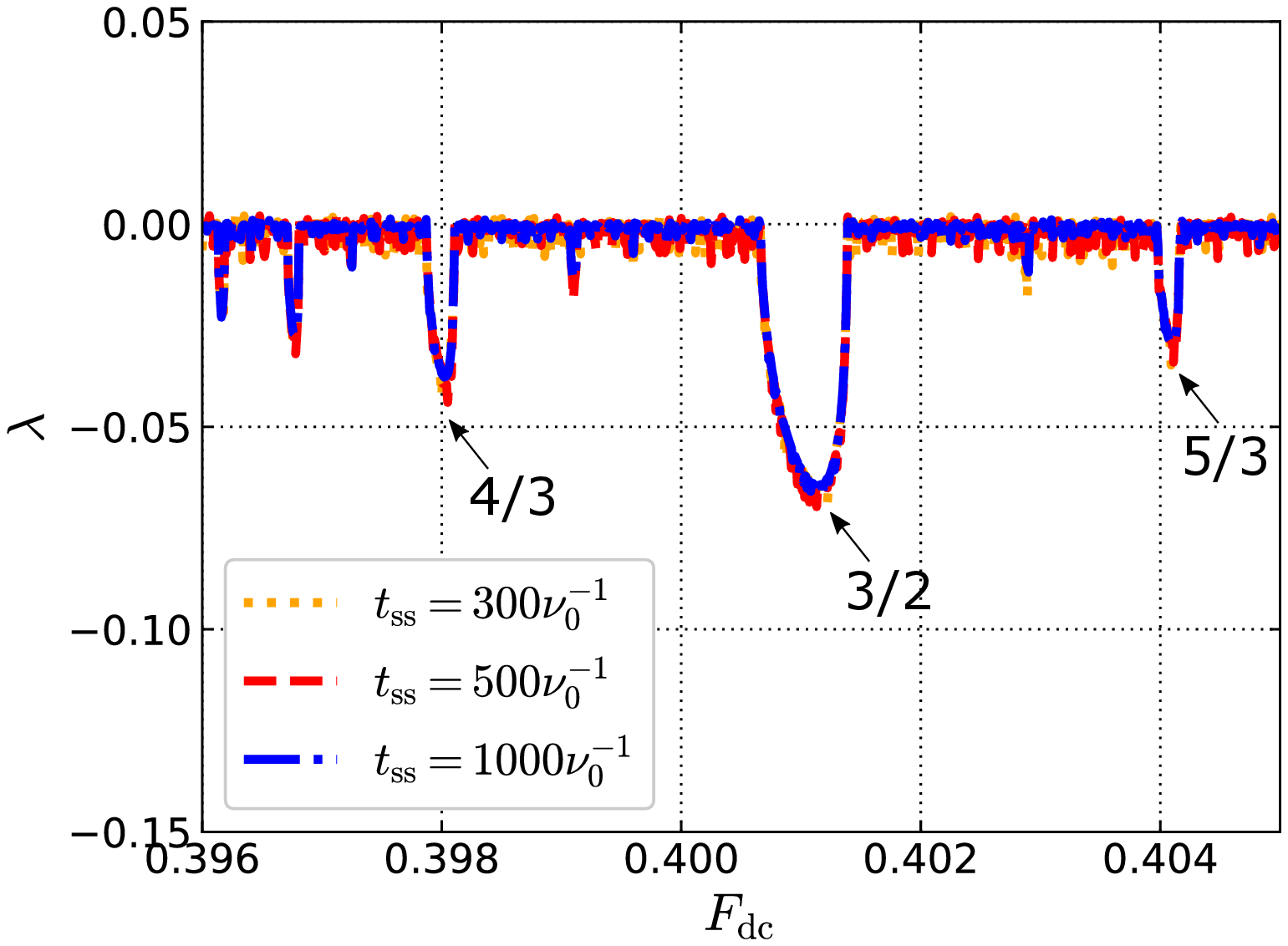}
\centering
\caption{Zoomed segment of Figure \ref{fig:f4} a). The largest LE is calculated for three values of $t_{\textrm{ss}}$. The numbers mark subharmonic steps.}
\label{fig:f5}
\end{figure}

For both quartic and exponential interparticle potential we briefly investigated the largest LE as a function of model parameters. The results have shown that the larges LE is always non-positive and thus the corresponding dynamics is non-chaotic. This is yet another property that is common for the family of overdamped FK models with different convex interparticle potentials. Also, just like in \cite{3}, the largest LE is negative in the pinned regime, reaching zero at the value $F_{\textrm{c}}$ for the first time. Afterwards the behaviour is consistent and the largest LE is negative for the entire interval of dc force where the corresponding Shapiro step is observed. In Figure \ref{fig:f5}, which represents a zoomed segment of Figure \ref{fig:f4} a), we observe that this tool is appropriate for detecting some subharmonic steps that could not be seen in the corresponding response function plot. One should note that small oscillations of the largest LE around $\lambda=0$ are consequence of using the finite-time $t_{\textrm{ss}}$ in numerical calculations. As can be seen from Figure \ref{fig:f5}, these oscillations get smaller as the time is increased and in the asymptotic limit $t_{\textrm{ss}}\to\infty$ they will completely vanish. 

\subsection{Bessel-like behaviour and Shapiro steps in the opposite direction of driving force}

\begin{figure}%[H]
\includegraphics[scale=0.4]{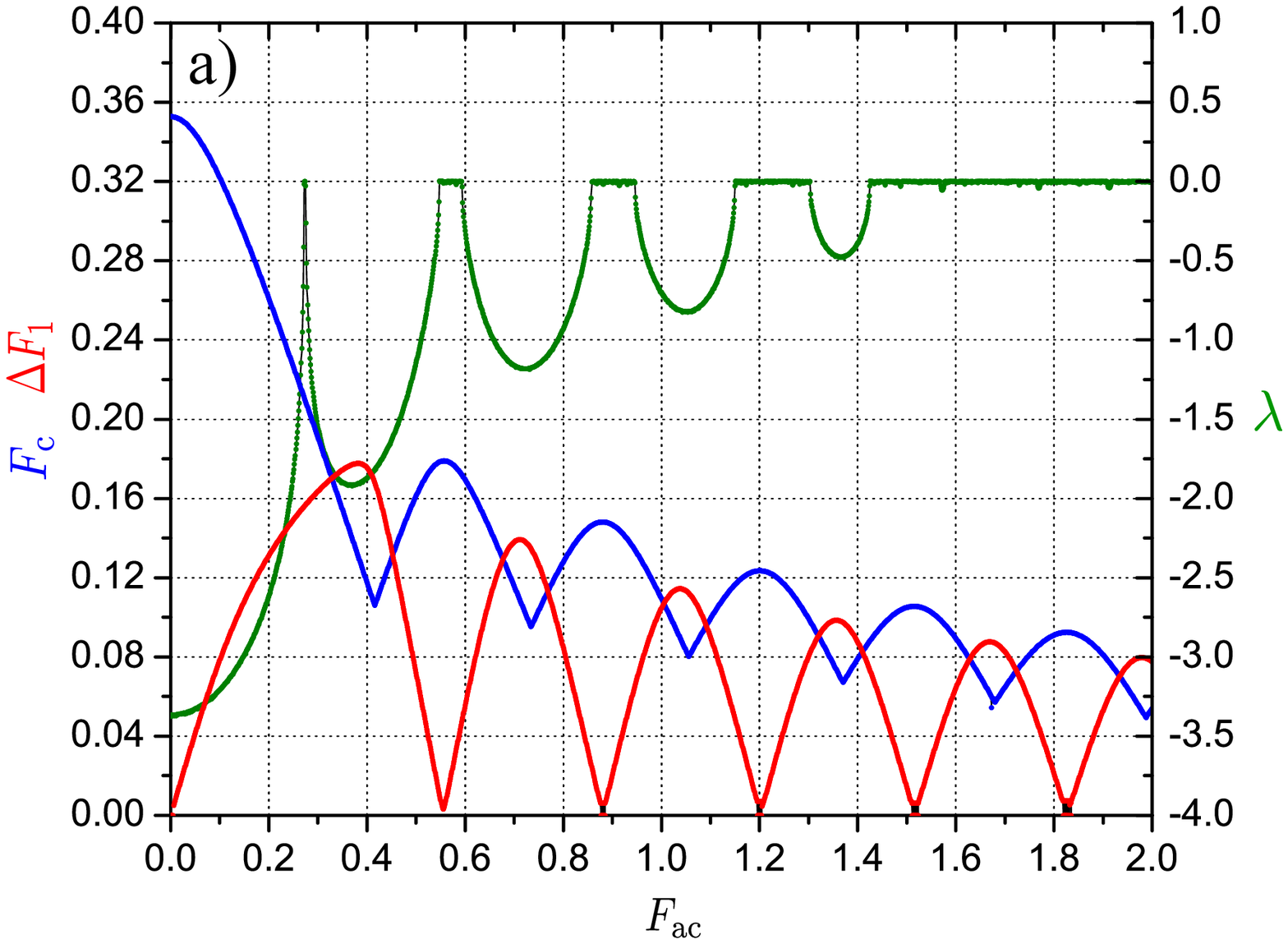} \hspace{2mm}
\includegraphics[scale=0.4]{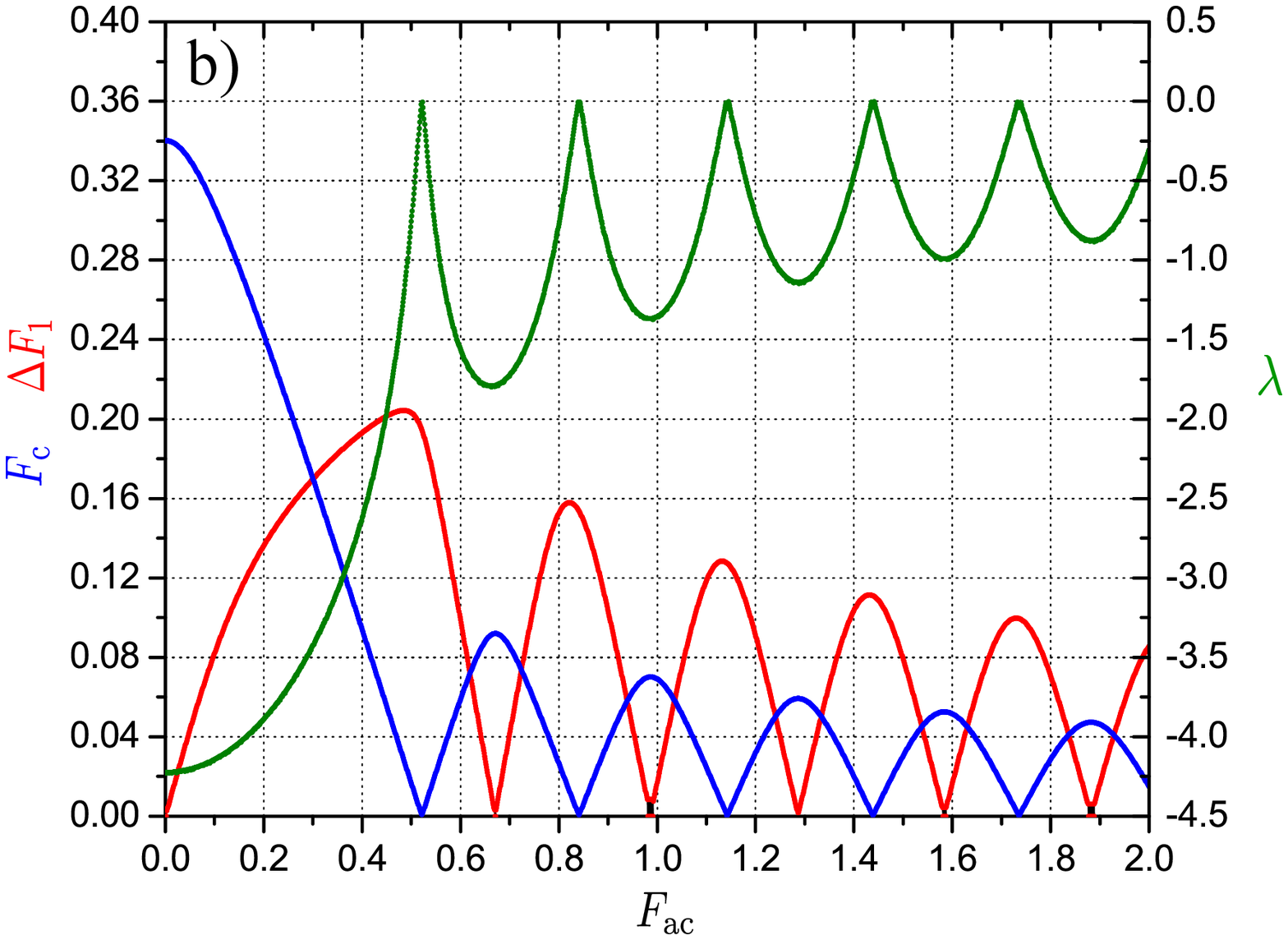}  
\centering
\caption{The critical force $F_{\textrm{c}}$, width of the first harmonic step $\Delta F_{1}$ and largest LE $\lambda$ for $F_{\textrm{dc}}=0$ as functions of the ac force amplitude $F_{\textrm{ac}}$ for two types of AIPs: a) quartic polynomial \eqref{5} and b) exponential interparticle potentials \eqref{6}. The chosen set of parameters is: $\omega=\frac{1}{2}$, $K=4.0$, $\nu_{0}=0.2$ and $g=h=f=1$.} 
\label{fig:f6}
\end{figure}

Reference \cite{3} provides evidence that, in the case of standard FK model, the largest LE can be used to investigate some characteristic properties of the model since the dependence of LE on $F_{\textrm{ac}}$ for $F_{\textrm{dc}}=0$ gives reverse image of the critical force amplitude dependence. Furthermore, amplitude dependencies of the critical force and first harmonic step width exhibit Bessel-like behaviour and the maxima of one function correspond to the minima of the other \cite{1,2}. However, different choice of the substrate potential in FK system leads to the violation of this rule \cite{3} and thus it would be interesting to examine whether the same would happen if some AIP form is taken into account. Amplitude dependencies of the critical force $F_{\textrm{c}}$, first harmonic Shapiro step width $\Delta F_{1}$ and largest LE $\lambda$ for $F_{\textrm{dc}}=0$ for the FK model with the quartic and exponential interparticle potentials are shown in Figure \ref{fig:f6}.

It is obvious that amplitude dependence of the critical force and width of the first harmonic step preserves the previously described behaviour in the standard case. This serves as yet another proof of the correspondence between the FK models with different convex interparticle potentials. However, the reverse image of largest LE and critical force is retained only for the exponential interparticle potential. Since there are some deviations in case of the quartic polynomial interparticle potential, this mirror image cannot be outlined as a general rule. 

\begin{figure}%[H]
\includegraphics[width=\columnwidth]{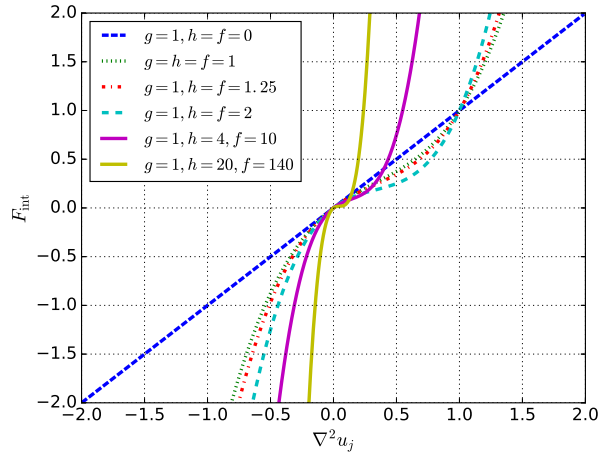}  
\centering
\caption{Interparticle force between the nearest neighbours from \eqref{*} ($F_{\textrm{int}}=g \nabla^2 u_j-h(\nabla^2 u_j)^2+f(\nabla^2 u_j)^{3}$) as a function of discrete Laplacian term $\nabla^2 u_j$ for several values of parameters $g$, $h$ and $f$.} 
\label{fig:f7}
\end{figure}

Yet another deviation from the standard model is observed in Figure \ref{fig:f6} a). While in the standard case it was shown that the value of critical depinning force is clearly equal to zero, or at least very close to it, for several values of the ac force amplitude that are in the same range as the one given in Figure \ref{fig:f6}, that is not the case when the quartic polynomial interparticle potential is taken into account. It becomes apparent that, in the case of AIP \eqref{5}, system has to overcome some additional force that tends to push the particles in the direction opposite to the direction of dc force. Simple analysis of the corresponding equations of motion \eqref{*} provides a credible explanation. Indeed, by plotting the interparticle force between the nearest neighbours $F_{\textrm{int}}=g \nabla^2 u_j-h(\nabla^2 u_j)^2+f(\nabla^2 u_j)^{3}$ as a function of discrete Laplacian term $\nabla^2 u_j$, as it is done in Figure \ref{fig:f7}, it becomes straightforward. Note that the negative values of $F_{\textrm{int}}$ present the situation when the direction of $F_{\textrm{int}}$ is opposite to the direction of dc force $F_{\textrm{dc}}$.  

In the purely harmonic case ($g=1$, $h=f=0$), interparticle force $F_{\textrm{int}}$ is obviously an odd function of discrete Laplacian term $\nabla^2 u_j$. In this case, critical depinning force has the smallest possible value (close to $F_{\textrm{dc}}=1.6$). However, in the case of non-zero values of parameters $h$ and $f$, the interparticle potentials are anharmonic and it is clearly seen that the influence of this term is biased towards the negative values of interparticle force. 

\begin{figure}%[H]
\includegraphics[scale=0.34]{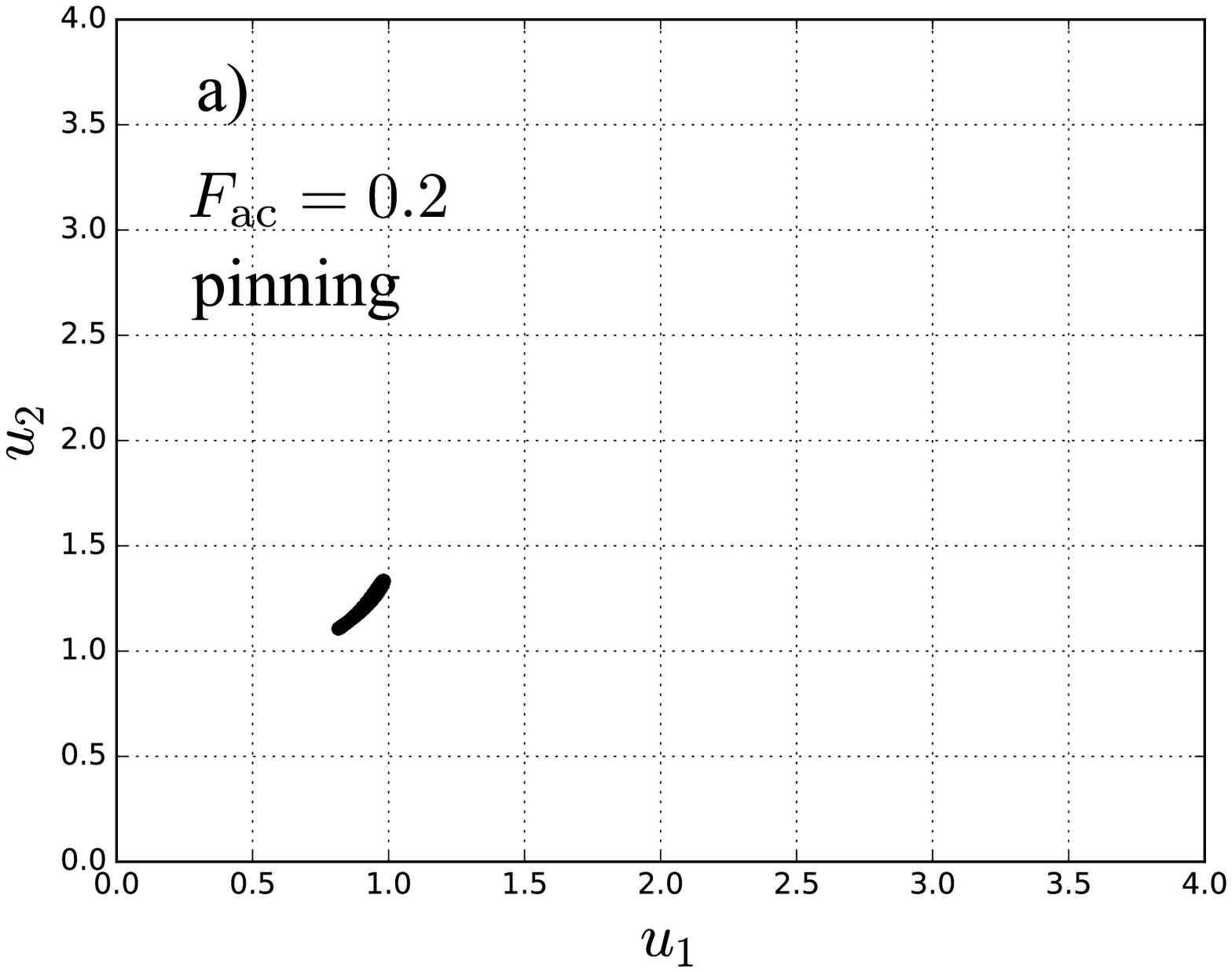}
\includegraphics[scale=0.34]{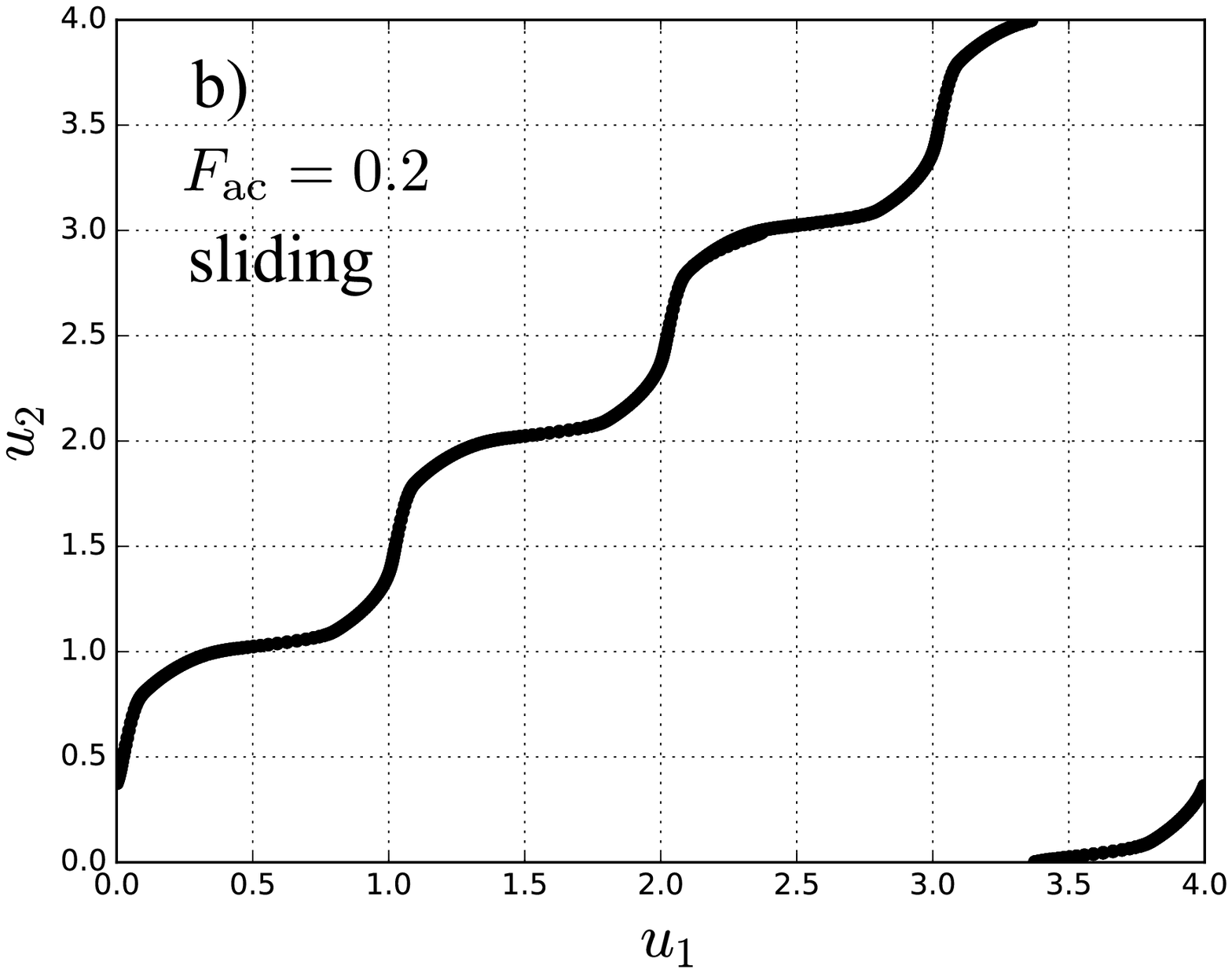}
\centering
\\
\includegraphics[scale=0.34]{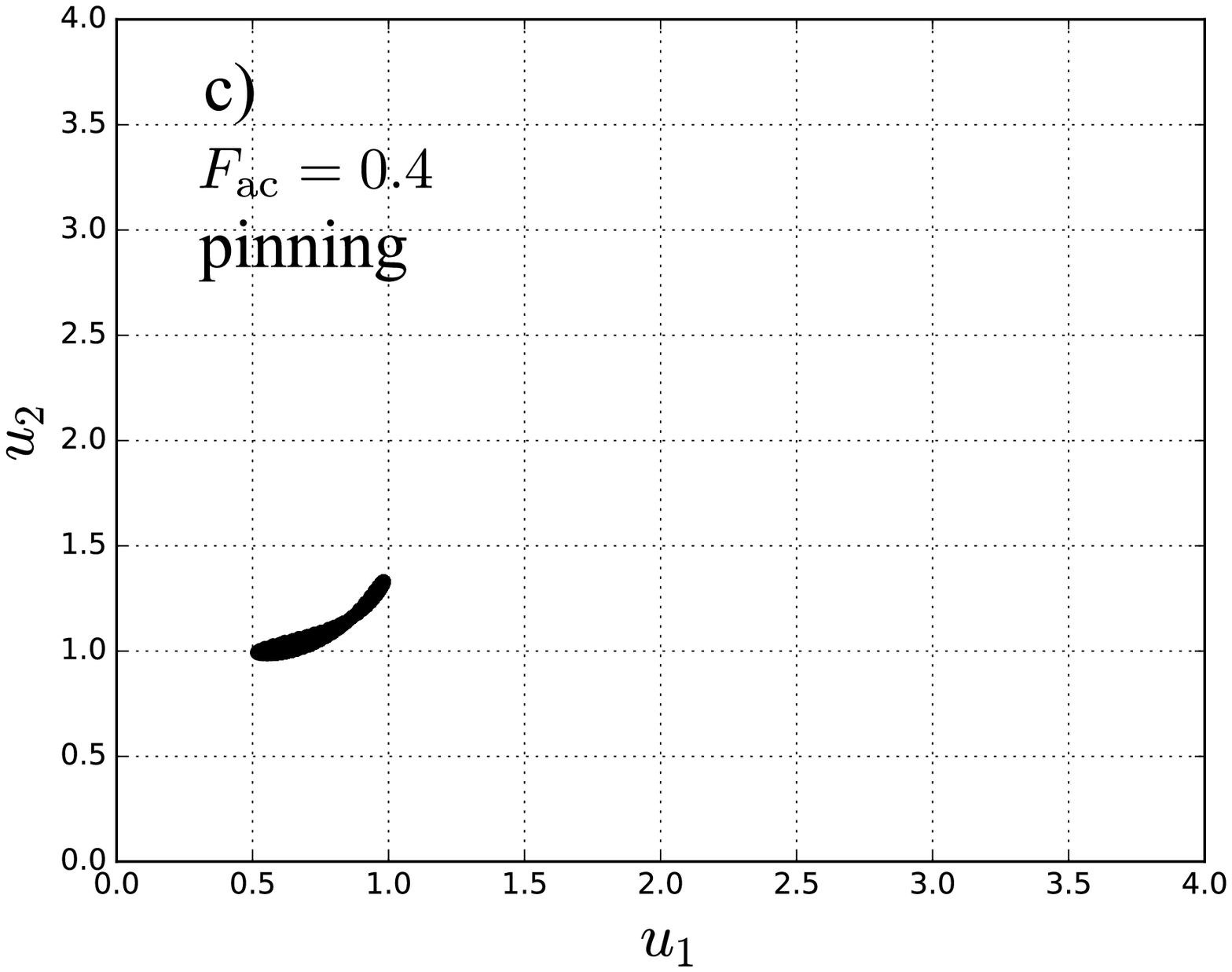}
\includegraphics[scale=0.34]{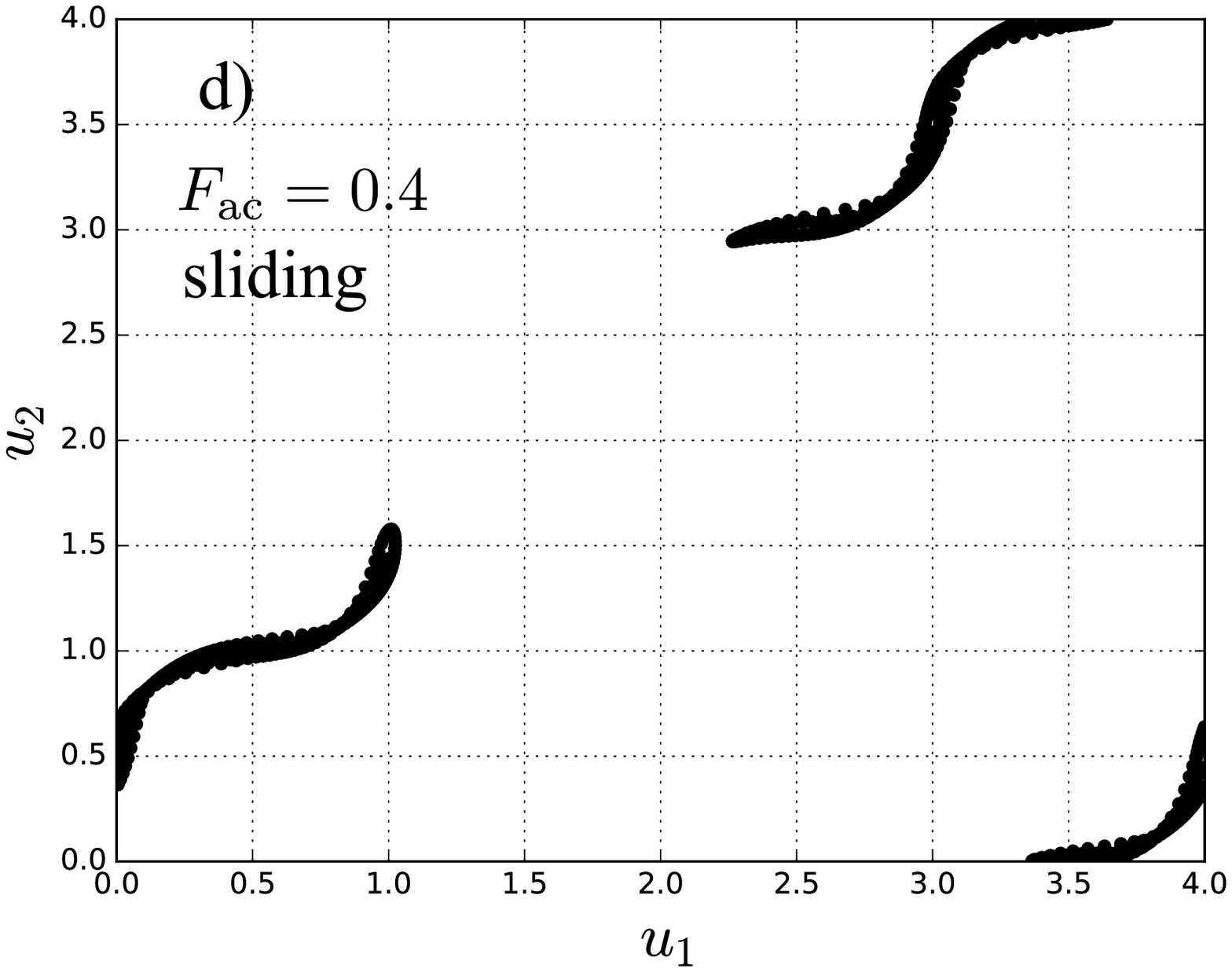}  
\centering
\caption{Poincar\'{e} sections for two neighbouring particles with coordinates $u_{1}$ and $u_{2}$ in the case of model with quartic polynomial interparticle potential \eqref{5} and chosen set of parameters: $\omega=\frac{1}{2}$, $K=4.0$, $\nu_{0}=0.2$ and $g=h=f=1$. The values of dc force are: a) $F_{\mathrm{dc}}=0.25$, b) $F_{\mathrm{dc}}=0.265$, c) $F_{\mathrm{dc}}=0.1$ and d) $F_{\mathrm{dc}}=0.12$.} 
\label{fig:f8}
\end{figure}

\begin{figure}%[H]
\includegraphics[scale=0.34]{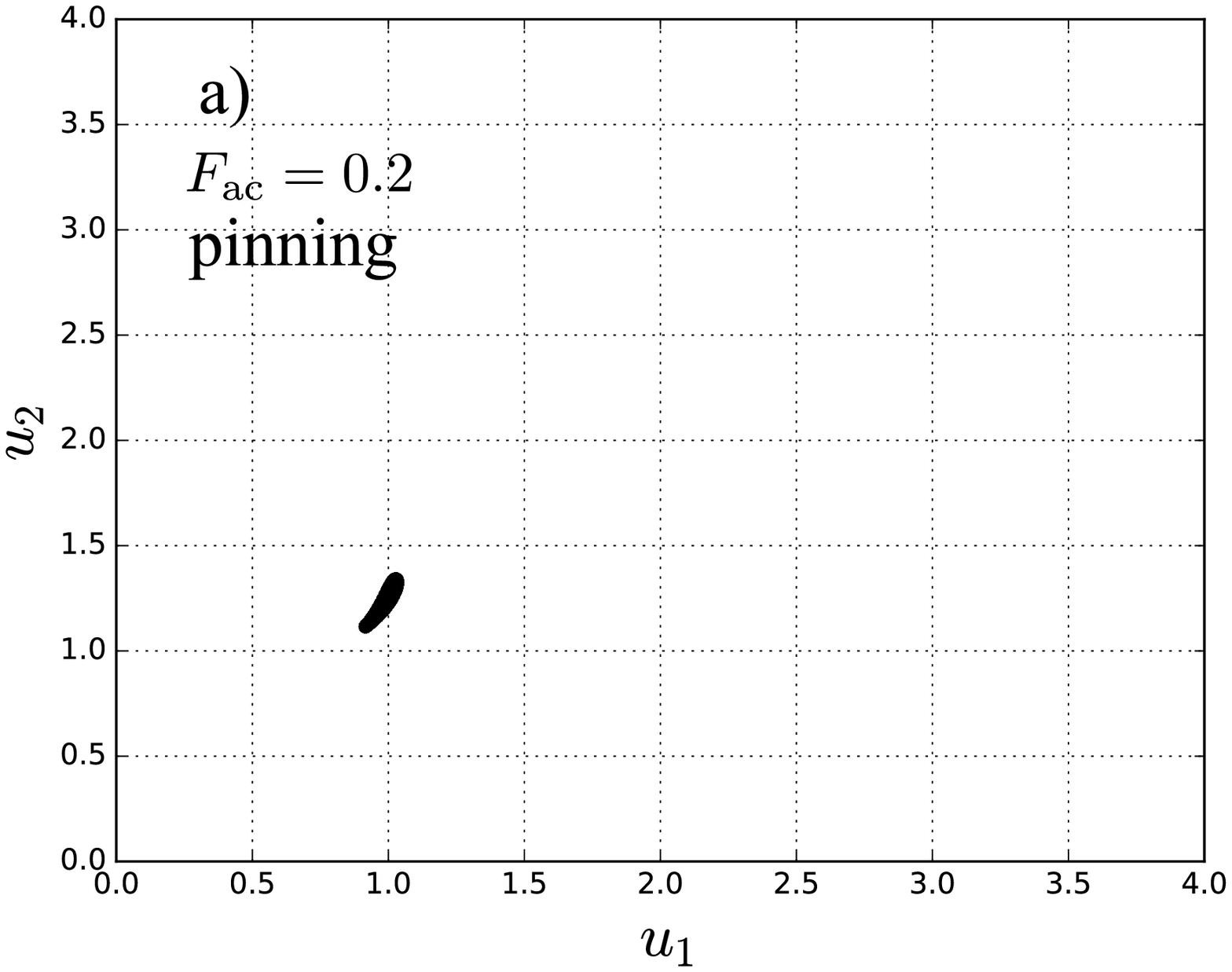}
\includegraphics[scale=0.34]{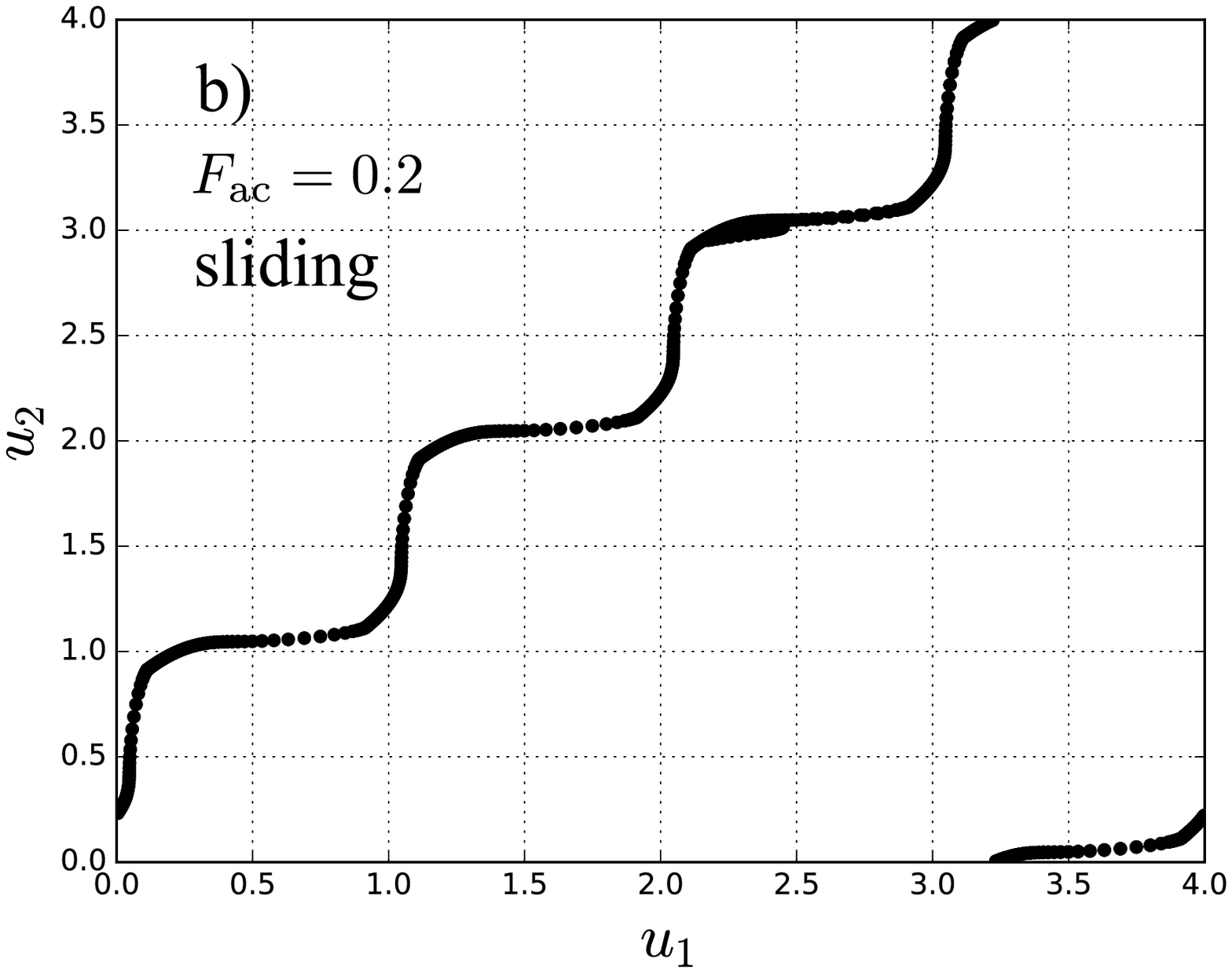}
\centering
\\
\includegraphics[scale=0.34]{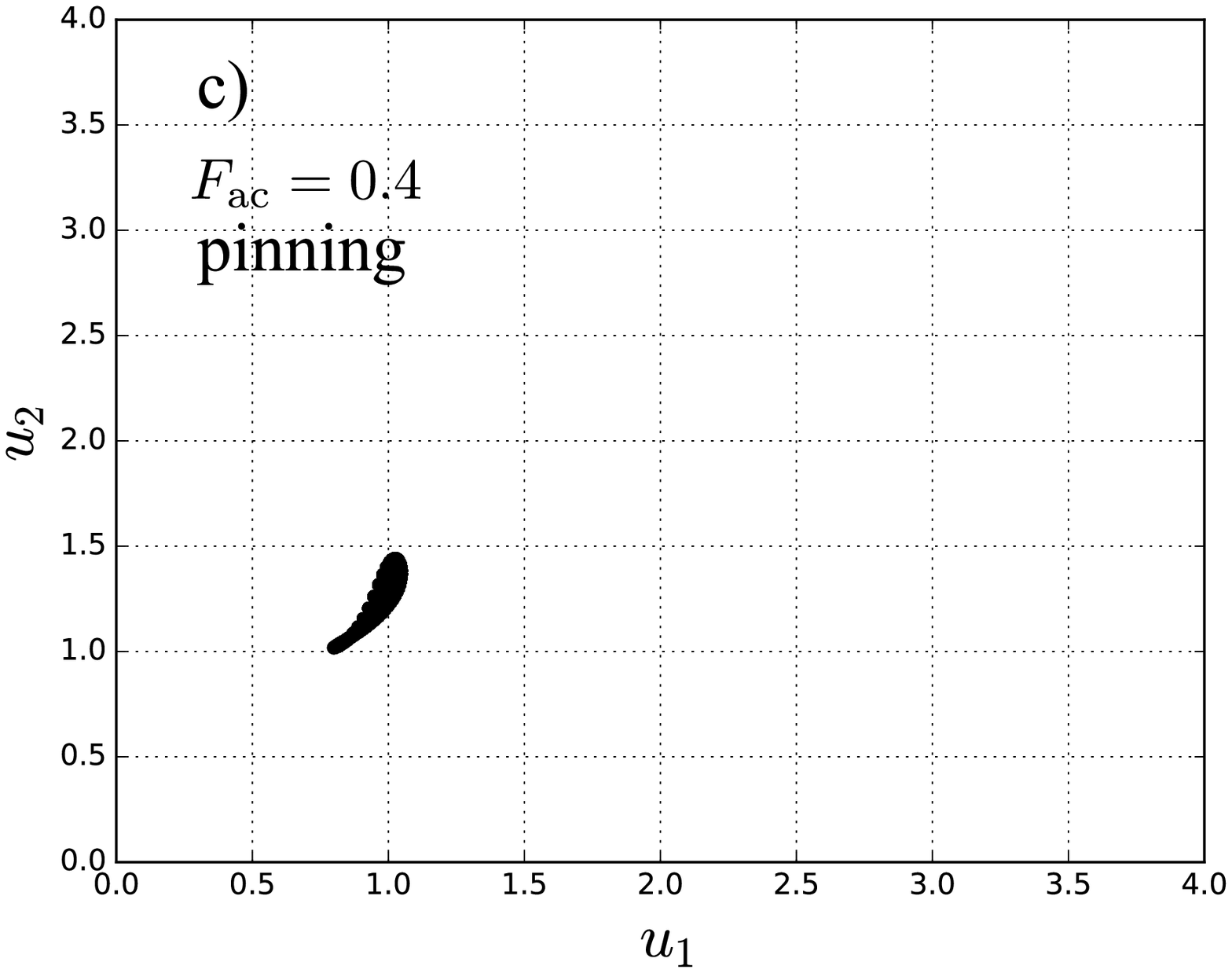}
\includegraphics[scale=0.34]{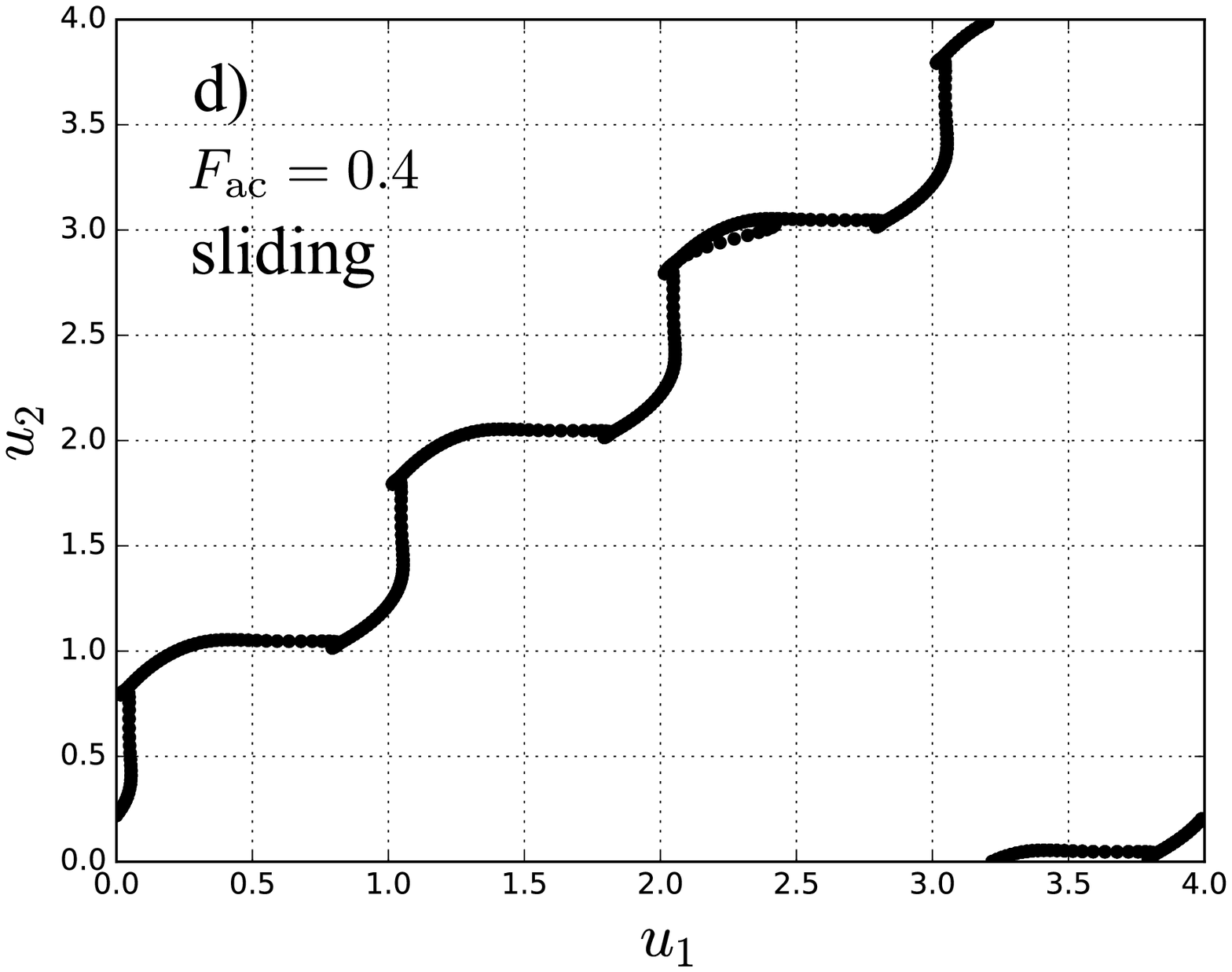}  
\centering
\caption{Poincar\'{e} sections for two neighbouring particles with coordinates $u_{1}$ and $u_{2}$ in the case of model with exponential interparticle potential \eqref{6} and chosen set of parameters: $\omega=\frac{1}{2}$, $K=4.0$ and $\nu_{0}=0.2$. The values of dc force are: a) $F_{\mathrm{dc}}=0.23$, b) $F_{\mathrm{dc}}=0.25$, c) $F_{\mathrm{dc}}=0.09$ and d) $F_{\mathrm{dc}}=0.1$.} 
\label{fig:f9}
\end{figure}

To check whether this deduction is correct and get a better insight into the particles' motion, Poincar\'{e} sections for two neighbouring particles with coordinates $u_{1}$ and $u_{2}$ are shown in Figures \ref{fig:f8} and \ref{fig:f9}. It is important to outline that, in order to be able to obtain such results, periodic boundary conditions were used and $N=8$ particles considered. Due to the fact that the period of substrate potential is $1$ and the winding number is $\omega=\frac{1}{2}$, the behaviour presented in these figures corresponds to the relative motion of two neighbouring particles in four potential wells. The results are presented for two values of ac force amplitude $F_{\mathrm{ac}}=0.2$ and $F_{\mathrm{ac}}=0.4$ and four dc force values that are close to the corresponding critical depinning forces in both pinning and sliding regime. While the positions of the particles are bound to a really tight band in the pinning regime, this is not the case with the sliding regime and the particles' collective motion is quite noticeable.

\begin{figure*}%[H]
\includegraphics[scale=0.65]{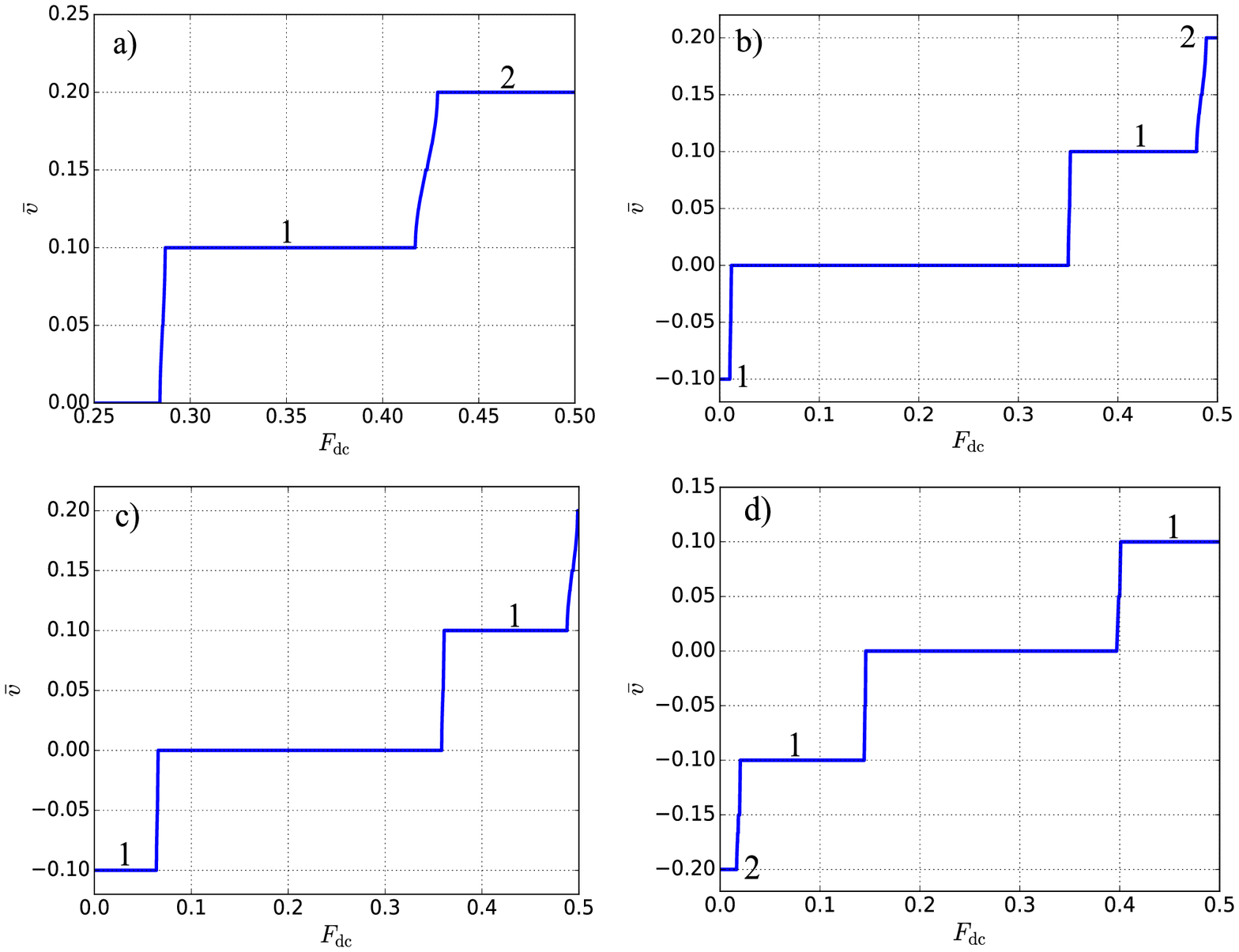} 
%\centering
\caption{Average velocity as a function of driving force in case of quartic polynomial interparticle potential \eqref{5} for $\omega=\frac{1}{2}$, $F_{\textrm{ac}}=0.2$, $K=4.0$, $\nu_{0}=0.2$ $g=1$ and several values of parameters $h$ and $f$: a) $h=f=1.25$, b) $h=f=2$, c) $h=4,f=10$ and d) $h=20,f=140$. The numbers mark harmonic steps.}
\label{fig:f10} 
\end{figure*}

Although discrete Laplacian term $\nabla^2 u_j$ is both positive and negative at different moments in time, the undoubted favouring of the interparticle force with direction opposite to the direction of driving dc force eventually gives rise to the critical depinning force $F_{\textrm{c}}$. This is rather an interesting observation and implies that the convenient choice of parameters $h$ and $f$ could cause reduction of the depinning force to zero value. This situation would occur in the case when the interparticle force term is strong enough to push the particles out of their pinned positions in the direction opposite to the direction of dc force. In Figure \ref{fig:f10} we present the evidence that this assumption is accurate.

In Figure \ref{fig:f10} a) interparticle force is still insufficient to push the particles out of their positions and the critical depinning force is a bit larger than it was the case for the set of parameters considered previously ($g=h=f=1$). This is purely the consequence of the fact that the biased direction of the interparticle force is the one opposite to the direction of dc force. However, with the increase in values of parameters $h$ and $f$, particles are pushed out of their positions in the direction opposite to the dc force and average velocity has a non-zero value even for $F_{\textrm{dc}}=0$. 

Of course, the negative sign of the average velocity is merely a consequence of the fact that its direction is opposite to the direction of dc force and is thus purely a statement of the convenient convention. Mode-locking is observed even in this region and thus the appearance of Shapiro steps is detected at the already expected values of average velocity: $\bar{v}=0.1$ (Figure \ref{fig:f10} b), c) and d)) and $\bar{v}=0.2$ (Figure \ref{fig:f10} d)) but now with a negative sign due to the opposite direction. When dc force becomes strong enough, average velocity decreases to zero. Afterwards, the behaviour is analogous to the one given in Figure \ref{fig:f1} and the dc force prevails over the interparticle force, leading the particles in the opposite direction. 

\section{Conclusion}

In this paper, generalized dissipatively driven FK models with anharmonic convex interparticle potentials are considered. The research was conducted for two types of interparticle potentials that resemble the ones that are already discussed in uniform atomic chains \cite{i1} and one-dimensional Hamiltonian systems perturbed by a conservative noise \cite{i2}. 

The response functions were obtained in the case of quartic polynomial \eqref{5} and exponential interparticle potentials \eqref{6}. Clear signs of correspondence between the FK models with different types of convex interparticle potentials were observed, but with two major differences: value of critical depinning force $F_{\textrm{c}}$ and size of Shapiro steps. The response functions for the model with the discussed anharmonic interparticle potentials match completely the results obtained in the standard case for integer values of winding number and the model reduces to a single particle model. However, the amplitude dependencies of the critical force $F_{\textrm{c}}$, size of the first harmonic step $\Delta F_{1}$ and largest LE $\lambda$ for $F_{\textrm{dc}}=0$ have shown that there are some deviations from the standard FK model if the considered interparticle potential is anharmonic. Although the mirror image of the amplitude dependence of critical depinning force and largest LE observed in the standard case holds out in case of the exponential interparticle potential, this conclusion cannot be drawn generally since in case of the quartic polynomial interparticle potential this image is not retained. We have shown that the reason for non-zero minima of the critical depinning force in case of the quartic polynomial interparticle potential lies in the fact that anharmonic interparticle potentials tend to push the particles in the direction opposite to the direction of the dc drive. However, when the dc force becomes large enough, it overcomes the influence of the interparticle force and the particles are pushed in the direction of the dc force. Another interesting result is that mode-locking phenomenon is observed in both directions of particles' motion. Namely, if one chooses interparticle forces that are strong enough to solely push the particles out of their pinned positions and applies external dc drive in the opposite direction, Shapiro steps are detected in the response function plots at both positive and negative values of average velocities. These values are determined by the equation \eqref{8} and the sign of negative average velocities is merely a consequence of the reverse direction of particles' motion. 

The FK models with anharmonic interparticle potentials have not been widely studied so far \cite{anharm,anharm2,shuk,brakiv}. In the present paper, it is shown that it is possible to change the direction of motion by conveniently choosing an anharmonic convex interparticle potential and vary the critical depinning force value and the size of Shapiro steps. Although it was already known that the critical depinning force decreases to zero for certain values of ac force amplitude in the standard FK model, this paper has provided evidence that this can also be done by choosing a convenient anharmonic form. Furthermore, critical depinning force non-zero minima have already been observed in experiments \cite{e3}. Therefore, our results could be significant for such systems as we have shown that anharmonicity might be the cause of it. The presented results could be important for the study of overdamped systems like irradiated Josephson junction arrays, where capacitance of junctions is small enough \cite{4}, colloidal systems \cite{e1,e2,c1,c2,c3} and charge-density wave systems \cite{thorn1,e3,thorn} since pure harmonicity is not so common in nature. Moreover, presented results can also be relevant in layered FK system \cite{TJL,TJL2}. 

\section*{Acknowledgment}

This work was supported by the Serbian Ministry of Education, Science and Technological Development of the Republic of Serbia under Contracts No. 171009. and No. III-45010 and by the Provincial Secretariat for High Education and Scientific Research of Vojvodina (Project No. APV 114-451-2201). This work was also supported by the Ministry of Education, Science and Technological Development of the Republic of Serbia (Grant No. 451-03-68/2020-14/200125).
         
%\section*{References}
\renewcommand\refname{}


\begin{thebibliography}{0}
\bibitem{fut1} 
Critical depinning force separates pinning from sliding regime. In other words, for small values of dc force particles of the system are pinned to the static configurations with zero average velocity (pinning regime) and above the threshold they have a non-zero average velocity (sliding regime).
\end{thebibliography}
\end{document}